\documentclass[preprint]{JHEP3} 

\usepackage{epsfig,multicol,bbm}

\newcommand\fverb{\setbox\pippobox=\hbox\bgroup\verb}
\newcommand\fverbdo{\egroup\medskip\noindent%
			\fbox{\unhbox\pippobox}\ }
\newcommand\fverbit{\egroup\item[\fbox{\unhbox\pippobox}]}
\newbox\pippobox

\title{Quantum construction of a unitary SU(2/1) model of the electro-weak interactions
with 2 Higgs doublets.}

\author{Jean Thierry-Mieg\\
NCBI. NLM. NIH Bldg 38A,\\
8600 Rockville Pike,\\
Bethesda, MD 20894, USA.\\
Tel   1 (301) 435 49 21   Fax   1 (301) 480 92 41
	E-mail: \email{mieg@ncbi.nlm.nih.gov}\\
and\\ 
Laboratoire de Physique Th\'eorique et d'Astroparticules,\\
CNRS, Montpellier, France.}
\abstract{The interactions and even the number of the Higgs
scalar fields are not fixed in the $SU(2)U(1)$ 
standard model of the electro-weak interactions
and the intrinsically chiral nature of the weak interactions is
not explained. Embedding $SU(2)U(1)$ into the Lie superalgebra $SU(2/1)$
fills these gaps.
The 2 smallest representations of $SU(2/1)$ adequately describe the electron, 
neutrino, up and down quarks and correlate their chiralities
with their $U(1)$ charges, and the Higgs fields have the
quantum numbers of the odd generators. But so far, there
was an apparent conflict with unitarity, because the
quark representation is not Hermitian and the super-Killing
metric is not positive definite. We solve this paradox
by assuming the existence of
2 complex Higgs doublets minimally 
coupled to the Fermions via the chiral projections of the odd 
generators of $SU(2/1)$. We find that Lagrangian induced by the Fermion loops
is unitary, thanks to the balance between the leptons and the
quarks needed to cancel the triangle anomaly and that the super-Jacobi
identity guarantees that the photon remains massless after symmetry breaking.
In addition, the Lagrangian has a  classical geometric interpretation in terms
of the curvature of the corresponding Hermitian algebra.
Assuming that the relative strength
of the scalar and vector couplings does not depend on the number of
families constrains the mass of the Higgs to 
$M^2_{H^0_1} + M^2_{H^0_2} = 32/9\;M^2_W = 2\;(107.2 GeV)^2$.
Contrary to grand-unified or Wess-Zumino supersymmetric models, the
$SU(2/1)$ internal superunification does not predict any unobserved 
particle besides the 2 Higgs doublets. 
}

\preprint{PTA/07-058}

\newcommand{\BE}{\begin{equation}}
\newcommand{\EE}{\end{equation}}
\newcommand{\BQ}{\begin{equation} \begin{array}{c}}
\newcommand{\EQ}{\end{array}\end{equation}}
\newcommand{\BT}{\begin{theorem}}
\newcommand{\ET}{\end{theorem}}

\newcommand{\bc}{\begin{center}}
\newcommand{\ec}{\end{center}}

\newcommand{\FHH}{\widehat{F}}
\newcommand{\gHH}{\widehat{g}}
\newcommand{\GT}{\widetilde{G}}

\newcommand{\AT}{\widetilde{A}}
\newcommand{\DT}{\widetilde{D}}
\newcommand{\FT}{\widetilde{F}}

\newcommand{\GX}{\Gamma}

\newcommand{\ALG}{\mathcal{A}}
\newcommand{\LAG}{\mathcal{L}}

\newcommand{\demi}{{1 \over 2}}

\begin{document} 


\section{Background}

The Weinberg-Salam $SU(2) . U(1)$ model of the
electro-magnetic and weak interactions \cite{Wei67,Sa68} is extremely accurate. At high
energy, the Fermions are described as massless $SU(2)$ doublets of
left spinors: the electron neutrino left doublet $(e_L,\nu _L)$ with
$U(1)$ charge (-1), and the up and down left quark doublet $(u_L,
d_L)$ with charge (2/3), accompanied by right singlets: the right
electron $(e_R)$ with charge (-2) and the right quarks $(u_R),
(d_R)$ with respective charges (4/3) and (-1/3). The right neutrino
is $SU(2)$ and $U(1)$ neutral and drops out from the model.

The corresponding $SU(2)$ and $U(1)$ Yang-Mills gauge fields \cite{YM54} are the
$(W^+,W^-)$, the $Z^0$ and the photon. The existence and
interactions of the $W^+$, the $W^-$ and the $Z^0$ were beautifully confirmed over
the last 30 years. The only component of the model that has not yet
been observed is the scalar Higgs field. In the minimal approach, it is
predicted to be a scalar $SU(2)$ doublet $\phi$ with $U(1)$ charge
(-1). The Higgs potential is assumed to be of the form
 $ V(\phi) = (\phi^2 - v^2)^2 $ So, in the vacuum the Higgs field
does not vanish. The gauge group breaks down to the electro-magnetic $U(1)$, the maximal subgroup
leaving $v$ invariant. The photon and the left neutrino remain massless. All other
 particles acquire a mass proportional to $v$ (see for example \cite{AL73}).

This construction is superb, but the number of arbitrary parameters is very large.
Since the left and right chiral states are independent, we need to choose 10
charges to describe the leptons and the quarks. For example, a massless 
electrically charged particle, which experimentally does not exist, would 
be acceptable at the classical level and is only ruled out by the study of 
the quantum anomalies. 
Also, since the Higgs field plays such a central role, one
would wish to derive it from first principle. The Yang-Mills gauge
fields are well understood, but the Higgs potential and even the number
of Higgs fields needed in the standard model remain arbitrary.

In 1979, Ne'eman \cite{N79} and Fairlie \cite{F79} have independently proposed
to consider the $SU(2/1)$ Lie superalgebra as a basis for describing
the weak interactions. Indeed, the fundamental representation, recalled in section 3,
exactly corresponds to the lepton triplet if we grade it by chirality.
The representation is Hermitian (3.2) and scaling the trace of all the even
matrices to a common value seemed to predict a electro-weak angle $sin^2 \theta_W = 1/4$,
a good value at the time, although higher than today's experimental value $.22$.
In addition, the odd generators have the quantum numbers of the Higgs fields.
The quarks were either left out \cite{F79} or mentioned as a counterargument \cite{N79}.
But remarkably, it was found a few weeks later \cite{DJ79,NT80} that $SU(2/1)$  
also admits a 4 dimensional representation, recalled in section 4, exactly
fitting the quarks, with 2 right singlets and a left doublet with electric charges $2/3$ and $-1/3$. 

These results, summarized in a recent Physics Report \cite{NSF05}, raised some interest
but also some doubts \cite{S92}. A superalgebraic structure seems in direct conflict
with unitarity. The supertrace, which is the natural invariant of a superalgebra, 
yields a negative sign for the propagator of the $U(1)$ vector, the
quark representation (4.3) is neither Hermitian nor anti-Hermitian, 
and even the Hermitian nature of the lepton representation is artefactual,
since the odd generators of antilepton representation are anti-Hermitian (3.8). It was therefore
often argued that the $SU(2/1)$ superalgebra could not play a role in
the quantum field theory. The present paper reverses the situation.

\section {Results}

By following literally the paradigm of the minimal couplings and turning
immediately to the renormalization theory, we predict 2 Higgs doublets $H$ and $K$.
This solves the problem of the metric
and allows to adjust freely the electro-weak angle. We prove that in the
symmetry-breaking vacuum the photon remains massless because of the
super-Jacobi identity. For a particular choice of the scalar coupling
constants, we find a scalar Ward identity insuring  that the 1-loop
counterterms do not depend on the number of families and derive
the mass relation
 $M^2_{H^0_1} + M^2_{H^0_2} = 32/9\;M^2_W = 2\;(107.2\;GeV)^2$.
The whole construction plays on balance between
the leptons and the quarks which ensures the cancellation
of the triangle anomaly \cite{BIM72}. The theory is however incomplete, because the 
scalar loops and the gluon loops
disturb the superalgebraic structure.

We start in section 5 from the assumption that the adjoint representation of the $SU(2/1)$ 
superalgebra directly describes the Bosons of the theory, and that the
quark and lepton representations, graded by chirality, describe their
interactions to the Fermions. The even generators are gauged as usual
as Yang-Mills vector fields. But since the odd generators are not Hermitian, 
we propose (5.3) to separate the left and right
interactions and construct 2 sets of scalar fields, $\Phi_L$ and
$\Phi_R$ by projecting out the relevant chiral part of the odd matrices,
and to define the interactions 
${\overline \Psi_L}\;\Phi_L\;\Psi_R + {\overline \Psi_R}\;\Phi_R\;\Psi_L$.
We then construct the rest of the Lagrangian by studying the counterterms
induced by the Fermion loops. 

As shown in (5.4), the left-trace of the odd matrices induces the propagator of the
scalar fields. It is antisymmetric and proportional to the super-Killing
metric, as befits a superalgebra. But since the $\Phi_L$ and
$\Phi_R$ are distinct, the propagator does not vanish by symmetrization and can be rediagonalized
(5.9) in the form of 2 Higgs scalar complex doublets, $H$ and $K$.
 $H$ is coupled to the negatively charged right
particles (electron and down quark) and $K$ to the positively charged
right up quark. This re-shuffling depends on the same balance between the leptons
and the quarks which cancels the triangle anomaly and the equations
are similar. In the later case, one computes the triple vector
Fermion loop, which is possibly spoiled by the impossibility to regularize the
chiral Fermions in a gauge invariant way, and depends on a supertrace condition (4.5). 
In the present paper we compute the 2 scalars and 4 scalars Fermion loops, which are a priori spoiled
by the non Hermitian nature of the $SU(2/1)$ scalar-Fermion couplings. In
all cases, we find that the respective contributions of the leptons and the quarks
cancel each other and induce a canonical counterterm.

In section 6, we construct the Higgs potential. A priori,
since we have 2 Higgs doublets, there are 3 possible $SU(2)U(1)$ invariant
quartic terms $V_1 =(H^2)^2 + (K^2)^2$ which controls the mass of
the neutral Higgs fields, a kind of vector product squared
$V_2(H,K)$ which controls the mass of the charged Higgs fields,
and a scalar product squared $V_3 = (H.K)^2$. 
We show that $V_2$ vanishes if $H$ and $K$ are parallel 
 as a consequence of the super-Jacobi identity,	
and that $V_3$ is absent, because of the lepton quark compensation. 
The potential can then
be rewritten in terms of  a simple quadratic form $\GT$ (6.2) which plays the
role of the Killing metric and contains a free parameter $\alpha$ later 
associated (10.3) to the electro-weak angle. This construction gives the
structure of the scalar potential, but not the strength of the couplings.

In section 7, we show that for a particular value of the scalar Fermion coupling, we
have a Ward identity insuring that the Fermion loops do not
modify the relative strength of the scalar and vector couplings
and that the scalar potential renormalizes like $g^2$. This
indicates that the mass of the neutral Higgs fields
should satisfy the relation $M^2_{H^0_1} + M^2_{H^0_2} = 32/9\;M^2_W = (151.6\;GeV)^2$.
In the symmetric situation, we would have $M_{H^0_1} = M_{H^0_2} = 107.2\;GeV$.
However, we also observe that other quantum corrections disrupt the 
universality of the scalar-Fermion coupling, indicating
that the theory is still incomplete.

In section 9, we show that the chiral decomposition of the Higgs
scalars, in conjunction with the Dirac 1-forms $\GX$ of \cite{TM06},
naturally constructs a covariant differential corresponding to the SU(2/1) 
Hermitian algebra first defined by Sternberg and Wolf \cite{SW78}.
The chirality operator and the complex Higgs structure conspire
so that the corresponding curvature 2-forms is well defined and
indeed valued in the adjoint representation of the $SU(2)U(1)$ even Lie algebra.

In section 10, we use the quadratic form (6.2) and the 2 CP conjugated
curvatures (9.12-13) to construct the vector Lagrangian
and show that the square of the H-algebra curvature tensor reproduces
the Higgs quartic potential induced by renormalization.
Within this $SU(2/1)$ framework, the electro-weak angle is free.

The theory is not complete and several important problems are presented
in the discussion. But even considering these difficulties, the results so far
are new and unexpected. Close to 30 years after the initial proposal \cite{N79,F79},
this is the first time that $SU(2/1)$ is shown to play a role in the
quantized theory, and the first time that the super-Killing metric
is used, both in the construction of the propagators (5.9 and 10.1) and of the
potential (6.2,6.4), while explicitly respecting unitarity.

\section{The $SU(2/1)$ lepton representation}

The smallest nontrivial simple Lie superalgebra is $SU(2/1)$, also
called $A(1/0)$ in the Cartan-Kac classification. The Lie sub-algebra
$\ALG_0$ is $SU(2) U(1)$, and the odd $\ALG_0$ module $\ALG_1$ is a
complex doublet with $U(1)$ charge $-1$. The fundamental representation
\cite{SNR77}
is of superdimension (2/1) and fits the leptons \cite{N79,F79}. 
In the $(\nu_L,\;e_L;\;e_R)$ basis, the 4 even generators read:
\BE
 \lambda_1 = \pmatrix { 0 & 1 & 0 \cr 1 & 0 & 0 \cr 0 & 0 & 0  }\,,\;\;
 \lambda_2 = \pmatrix {  0 & -i & 0 \cr i & 0 & 0 \cr 0 & 0 & 0  }\,,\;\;
 \lambda_3 =  \pmatrix { 1 & 0 & 0 \cr 0 & -1 & 0 \cr 0 & 0 & 0  }\,,\;\;
 \lambda_8 =   \pmatrix { 1 & 0 & 0 \cr 0 & 1 & 0 \cr 0 & 0 & 2  } \,.
\EE
the 4 odd generators read:
\BE
 \lambda_4 =  \pmatrix { 0 & 0 & 1 \cr 0 & 0 & 0 \cr 1 & 0 & 0  }\,,\;\;
 \lambda_5 =  \pmatrix { 0 & 0 & -i \cr 0 & 0 & 0 \cr i & 0 & 0  }\,,\;\;
 \lambda_6 =  \pmatrix { 0 & 0 & 0 \cr 0 & 0 & 1 \cr 0 & 1 & 0  }\,,\;\;
 \lambda_7 =  \pmatrix { 0 & 0 & 0 \cr 0 & 0 & -i \cr 0 & i & 0  }\,,
\EE
and the chirality operator is
\BQ
\chi = 	diag(1,1,-1)\,.
\EQ 
By inspection, we find that the anticommutator of the odd matrices close on the
even matrices, defining the symmetric structure constants $d^a_{ij}$  and 
that the commutators of the odd matrices close on $\chi$ times the even matrices,
defining the skew-symmetric structure constants $f^a_{ij}$:
\BQ
 \lambda_i\, \lambda_j + \lambda_j\, \lambda_i = d^a_{ij}\,\lambda_a\;,
\\
 \lambda_i\, \lambda_j - \lambda_j\, \lambda_i = i\;\chi\;f^a_{ij}\,\lambda_a\;,
\\ 
a = 1,2,3,8\,\,;\;\;\;i,j = 4,5,6,7
\EQ
in these conventions, the $f^a_{ij}$ and $d^a_{ij}$ constants are real.
Finally, the vacuum $v$ is chosen along the $\lambda_6$ direction, whose
centralizer is the photon:
\BE
\hbox{photon} = - (\lambda_6)^2 = {1 \over 2}\; (\lambda_3 - \lambda_8) = \pmatrix { 0 & 0 & 0 \cr 0 &  -1 & 0 \cr 0 & 0 & -1 } \,.
\EE
There is no loss of generality in this choice of $v$, it
just corresponds to the way we named the particles. If we rotate
$v$ we just have to rotate the name of the electron and the neutrino
and keep the photon in the direction of $v^2$.
Notice that, with respect to the $SU(2/1)$ supermetric :
\BQ
Str (.) = Tr (\chi\; .)
\\
\gHH_{MN} = 1/2\; STr (\lambda_M \lambda_N)\;,\;\; M,N = 1,2...8
\\
 \gHH_{ab} = diag(1,1,1,-1)\,,\qquad \gHH_{45} = - \gHH_{54} = \gHH_{67} = - \gHH_{76} = i
\EQ
the photon is on the light-cone of the superalgebra.

The even generator of antilepton representation, in the $({\overline {(e_R)}}_L ; {\overline {(e_L)}}_R, {\overline {(\nu_L)}}_R)$ basis, are:
\BE
 \lambda_1 = \pmatrix { 0 & 0 & 0 \cr 0 & 0 & 1 \cr 0 & 1 & 0  }\,,\;\;
 \lambda_2 = \pmatrix { 0 & 0 & 0 \cr 0 & 0 & -i \cr 0 & i & 0 }\,,\;\;
 \lambda_3 = \pmatrix { 0 & 0 & 0 \cr 0 & 1 & 0 \cr 0 & 0 & -1 }\,,\;\;
 \lambda_8 = \pmatrix {-2 & 0 & 0 \cr 0 & -1 & 0 \cr 0 & 0 & -1  } \,.
\EE
the 4 odd generators:
\BE
 \lambda_4 =  \pmatrix { 0 & 0 & -1 \cr 0 & 0 & 0 \cr 1 & 0 & 0  }\,,\;\;
 \lambda_5 =  \pmatrix { 0 & 0 & i \cr 0 & 0 & 0 \cr i & 0 & 0  }\,,\;\;
 \lambda_6 =  \pmatrix { 0 & 1 & 0 \cr -1 & 0 & 0 \cr 0 & 0 & 0  }\,,\;\;
 \lambda_7 =  \pmatrix { 0 & -i & 0 \cr -i & 0 & 0 \cr 0 & 0 & 0  }\,,
\EE
and the chirality operator is
\BQ
\chi = 	diag(1,-1,-1)\,.
\EQ 
Equations (3.4-5) are again true in this representation. 
Notice however that the odd matrices (3.8) are anti-Hermitian, this is necessary to
maintain (3.5) since the electric charge of the antielectron is positive.
Notice also that the charge and parity (CP) are linked. The doublet and antidoublet must be of
opposite parity with respect to $\chi$ (3.3) and (3.9) in order to maintain the commutator (3.4b).
This switches the sign of the supermetric (3.6), as befits a CP invariant theory.

\section{The $SU(2/1)$ quark representation}

In the standard model \cite{Wei67,Sa68,GIM70,BIM72}, each family of quarks consists of 4 states,
a left doublet and two right singlets, i.e. $(u_R/(u_L,d_L)/d_R)$ for
the 'electron family'. 
Amazingly, the second smallest irreducible representation of the
$SU(2/1)$ superalgebra corresponds to this chiral decomposition, 
it contains a free parameter $n$ which allows us to fix the charge of the up quark.
From the mathematical point of view \cite{SNR77}, the existence of this 4 dimensional
representation is a simple consequence of the isomorphism between $SU(2/1)$ and $OSp(2/2)$,
which is a generalization of the well known isomorphisms between the first members
of the infinite families of simple Lie algebras. But from the physical point
of view, it came as a great surprise. In the original papers \cite{F79,N79},
the quarks were respectively left out and listed as a counterargument. The 
incorporation a few weeks later of the quarks \cite{DJ79,NT80} really appeared as a significant 
confirmation of the model.
The 4 even generators read:
\BE
 \lambda_1 = \pmatrix { 0 & 0 & 0 & 0 \cr 0 & 0 & 1 & 0 \cr 0 & 1 & 0 & 0 \cr 0 & 0 & 0 & 0  }\,,\;\;
 \lambda_2 = \pmatrix {  0 & 0 & 0 & 0 \cr 0 & 0 & -i & 0 \cr 0 & i & 0 & 0 \cr 0 & 0 & 0 & 0  }\,,\;\;
\EE
\BE
 \lambda_3 =  \pmatrix { 0 & 0 & 0 & 0 \cr 0 & 1 & 0 & 0 \cr 0 & 0 & -1 & 0 \cr 0 & 0 & 0 & 0  }\,,\;\;
 \lambda_8 =  {1 \over n} \; \pmatrix { -n-1 & 0 & 0 & 0 \cr 0 & -1 & 0 & 0 \cr 0 & 0 & -1 & 0 \cr 0 & 0 & 0 & n-1  } \,.
\EE
The 4 odd generators are not Hermitian, for any value of $n$, they read:
\BQ
 \lambda_4 =   { 1\over\sqrt {2n} }\pmatrix { 0 & 0 & -\sqrt {n+1} & 0 \cr 0 & 0 & 0 & \sqrt {n-1} \cr \sqrt {n+1} & 0& 0 & 0 \cr 0 & \sqrt{n-1} & 0 & 0  }\,,\;\;
\\ \\
 \lambda_5 =   { 1\over\sqrt {2n} }\pmatrix { 0 & 0 & i\sqrt {n+1} & 0\cr 0 & 0 & 0 & -i \sqrt {n-1} \cr i\sqrt {n+1} & 0 & 0 & 0 \cr 0 & i \sqrt {n-1} & 0 & 0  }\,,\;\;
\\ \\
 \lambda_6 =   { 1\over\sqrt {2n} }\pmatrix { 0 & \sqrt {n+1} & 0 & 0 \cr -\sqrt {n+1} & 0 & 0 & 0 \cr 0 & 0 & 0 & \sqrt {n-1}\cr 0 & 0 & \sqrt {n-1} & 0  }\,,\;\;
\\ \\
 \lambda_7 =   { 1\over\sqrt {2n} }\pmatrix { 0 & -i\sqrt {n+1} & 0 & 0 \cr -i\sqrt {n+1} & 0 & 0 & 0 \cr 0 & 0 & 0 & -i \sqrt {n-1} \cr 0 & 0 & i \sqrt {n-1} & 0  }\,.
\EQ
Finally the chirality operator is
\BQ
\chi = 	diag(-1,1,1,-1)\,.
\EQ
We have normalized the quark matrices (4.1-4.4) so that they have 
exactly the same supercommutators
as the lepton matrices (3.1,3.2). As desired,
the super-Killing metric (3.6)
is also identical in the lepton and the quark representation.
Notice however that the commutators of the odd quark matrices do not close on $\chi$ times the even matrices.

Since $n$ is a free parameter, $SU(2/1)$ does not predict the charge of the quarks, or the
number of colors. However, the superalgebra does relate the 
$U(1)$ charges of the singlets and a number of desired relations (post-dictions) are
automatically satisfied.

First, for $n=-1$ (or $n=1$), we recover the lepton (or antilepton) representation with only 3 states.
In other words, $SU(2/1)$ implies that if the charge of the electron is equal to
the charge of the $W$ vector, the $U(1)$ neutral right-neutrino is also 
neutral with respect to the odd generators.

Then, we should impose the cancellation of the triangle anomalies.
\BQ
  d_{abc} = STr_f (\lambda_a (\lambda_b \lambda_c + \lambda_c \lambda_b)) = 0\,,\;\;\;a,b,c = 1,2,3,8\;,
\EQ
If we consider a lepton family, with one lepton of $U(1)$ charge $1$, and $n'$ quarks of $U(1)$ charge $-1/n$,
the $SU(2)^2 U(1)$ anomaly vanishes if and only if we choose $n = n'$. But although there are
no other adjustable parameters, $SU(2/1)$ then implies that the $U(1)^3$ anomaly also vanishes
\BQ
STr_f \;(\lambda_8^3) = 2 - 8 + n\; ((n+1)^3 -2 - (n-1)^3)/n^3 = 0
\EQ
for any value of $n$. 
The vacuum and the photon direction were already chosen in the lepton representation. In the same 
conventions (3.5) where the electric charge of the electron is -1, we find:
\BE
\hbox{photon} = - (\lambda_6)^2 = {1 \over 2}\; (\lambda_3 - \lambda_8) = {1 \over 2n}\;\pmatrix { n+1 & 0 & 0 & 0 \cr 0 & n+1 & 0 & 0 \cr 0 & 0 & 1-n & 0 \cr 0 & 0 & 0 & 1-n  } \,.
\EE
Choosing the correct number of colors $n=3$ implies
the correct electric charges  for the $up$ quark $2/3$ and the $down$ quark $-1/3$ .
Finally, the odd sector antisymmetric structure constants and trace metric also vanish:
\BQ
   f_{aij} = Tr_f (\lambda_a (\lambda_i \lambda_j - \lambda_j \lambda_i)) = 0\,,\;\;\;a = 1,2,3,8\,,
\\
   g_{ij} = \demi Tr_f (\lambda_i \lambda_j) = 0\,,\;\;\;i,j = 4,5,6,7\,.
\EQ
We will now study in the next sections some consequences of these identities.

\section {The scalar propagator}

We would like to construct a quantum field theory, based on the $SU(2/1)$
superalgebra, which would extend the  Yang-Mills theory \cite{YM54}
associated to its maximal $SU(2)U(1)$ Lie sub-algebra and incorporate in
some way the odd matrices. In \cite{TM06}, we have shown that we
can construct an associative covariant exterior differential, mixing left
and right chiral Fermions, if and only if the Fermion form a representation
of a Lie superalgebra, graded by the chirality. As we have just seen, 
these conditions are met exactly 
by the existing fundamental Fermions: the leptons and the quarks.
The connection is then defined as the sum of the usual Yang-Mills 1-form 
$A_{\mu}\;dx^{\mu}$
plus  the constant Dirac 1-form $\gamma_{\mu}\;dx^{\mu}$ multiplied by scalar 
field valued in the odd sector of the superalgebra. 

The natural idea would therefore be to introduce a scalar-field $\Phi$ with a minimal
coupling to the Fermions of the form
$\Psi^{\dagger}\; \Phi^i \lambda_i \;\Psi $. However, this is not directly possible
since the $\lambda_i$ are not Hermitian. To fix this problem, we
postulate that the field $\Phi^i$ is composed of 2 parts, each with intrinsically
chiral couplings. Considering the chirality operator
$\chi$ (3.3,4.4) with eigenvalue $1$ $(-1)$ on the left (right) Fermions, 
we define the left and right projectors $p_L = (1 + \chi)/2$ and
$p_R = (1 - \chi)/2$, and define, for any matrix M, the chiral traces.
\BQ
Tr_L (M) = Tr (p_L\;M)\;,\;\;Tr_R (M) = Tr (p_R\;M)\;,\;\;Str (M) = Tr (\chi\;M)\;.
\EQ
 We then decompose the odd $\lambda$ matrices as
\BQ
  \lambda_i = \lambda_{iL} + \lambda_{iR}
\\
  \lambda_{iL} = p_L\;\lambda_{i} = \lambda_{i}\; p_R\;,\;\;
  \lambda_{iR} = p_R\;\lambda_{i} = \lambda_{i}\; p_L\;\;,
\EQ
and assume the existence of 2 scalar fields $\Phi_L$ which absorbs right Fermions
and emits left ones and $\Phi_R$ which absorbs left Fermions and emits right
ones, according to the coupling
\BQ
\Psi^{\dagger}_L\; \Phi^i_L \lambda_{iL} \;\Psi_R \; + 
\Psi^{\dagger}_R\; \Phi^i_R \lambda_{iR} \;\Psi_L \;.
\EQ
To find the correct propagator of the $\Phi$ field, we compute the quantum
corrections due to the insertion of a Fermion loop between 2 $\Phi$ states. 

This induces a scalar counterterm of the form
\BQ
\LAG_s =  \eta_{ij} \; \partial_{\mu} \Phi^i_L \; \partial_{\mu} \Phi^j_R\; 
\EQ
where
\BQ
\eta_{ij} =  {1 \over n+1} Tr_L (\lambda_i \; \lambda_j) =   {1 \over n+1} Tr_R (\lambda_j \; \lambda_i)\,.
\EQ
There are two contributions, one coming from the lepton loop and one from the quark loop
\BQ
(\eta_{ij})_{lepton} = {1 \over n+1} \delta_{ij} + {i \over n+1} \;\epsilon_{ij}\;,
\\
(\eta_{ij})_{quark} = - {1\over n(n+1)} \delta_{ij} + {i \over n+1} \;\epsilon_{ij}\,,
\EQ
where  $\epsilon_{45} = - \epsilon_{54} = \epsilon_{67} = - \epsilon_{76} = 1$, 
all other components being zero, which coincides with the
odd part of (3.6).
Summing over a whole family $f$ with $n$ colored quarks, we find that the Fermion 
loop is proportional to the canonical supermetric of $SU(2/1)$:
\BQ
(\eta_{ij})_f =  i\;\epsilon_{ij}\; =  \gHH_{ij}
\EQ

For the moment, we have a weird scalar model, with
an antisymmetric propagator and non Hermitian couplings. But now a miracle occurs. 
If we change variables as follows, the Lagrangian becomes canonical.
Consider the two $SU(2/1)$ odd valued scalar fields $H$ and $K$:
\BQ
\Phi_{iL} = a_i + i \epsilon_{ij} b_j\;,\;\;
a_i = \demi\; (H_i + i\;K_i)\;,\\
\\
\Phi_{iR} = a_i - i \epsilon_{ij} b_j\;,
\;\;b_i = \demi\; (H_i - i\;K_i)\;,
\EQ
where $a_i$ and $b_i$ are convenient intermediate variables which are not 
reused below.
Substituting (5.8) in (5.4) we obtain the canonical propagator
\BQ
\LAG_s = i \; \epsilon_{ij} \; \partial_{\mu} \Phi^i_L \; \partial_{\mu} \Phi^j_R\; 
= \demi  \delta_{ij} \; (\partial_{\mu} H^i \;  \partial_{\mu} H^j\;  +
\partial_{\mu} K^i \;  \partial_{\mu} K^j\;)\,.
\EQ
We then define the $\mu^{\pm}$ Hermitian matrices:
\BQ
   \lambda_i =  (\mu^-_i - i\; \mu^+_i)
\,,\qquad
   \mu^-_i =  {1 \over  2 } \; (\lambda_i  + \lambda_i^{\dagger})
\,,\\
   \lambda_i^{\dagger}  = (\mu^-_i + i\; \mu^+_i)
\,,\qquad
   \mu^+_i =  {i \over 2 } \; (\lambda_i  - \lambda_i^{\dagger})
\,,
\EQ 
and rewrite the scalar-Fermion interaction in terms of these new variables:
\BE
 \Psi^{\dagger}\;(\Phi^i_L \; \lambda_{iL} + \Phi^i_R \; \lambda_{iR})\;\Psi
     =   \Psi^{\dagger}\;(H^i \;\mu^-_i + K^i\;\mu^+_i )\;\Psi\,.
\EE
The matrices $\mu^{\pm}_i$ split according to the electric charge and are Hermitian. 
The field $H^i$ interacts only with the negatively
charged right singlets (electron, down quark) and the field $K^i$ just with
the positively charged right singlets (up quark). In the lepton
representation, the $\mu^-_i$
matrices are equal to the Hermitian matrices given in (3.2) 
\BQ
 \mu^-_4 =   \pmatrix { 0 & 0 & 1 \cr 0 & 0 & 0 \cr 1 & 0 & 0  }\,,\;\;
 \mu^-_5 =   \pmatrix { 0 & 0 & -i \cr 0 & 0 & 0 \cr i & 0 & 0  }\,,\;\;
\\
\\
 \mu^-_6 =   \pmatrix { 0 & 0 & 0 \cr 0 & 0 & 1 \cr 0 & 1 & 0  }\,,\;\;
 \mu^-_7 =  \pmatrix { 0 & 0 & 0 \cr 0 & 0 & -i \cr 0 & i & 0  }\,.
\EQ
In the quark representation, the $\mu^-_i$ are equal to the right and low Hermitian
corner of the matrices (4.3) 
\BQ
 \mu^-_4 =   \sqrt { n-1 \over 2n } \; \pmatrix { 0 & 0 & 0  & 0 \cr 0 & 0 & 0 & 1 \cr 0 & 0& 0 & 0 \cr 0 & 1 & 0 & 0  }\,,\;\;
 \mu^-_5 =   \sqrt { n-1 \over  2n } \; \pmatrix { 0 & 0 & 0  & 0\cr 0 & 0 & 0 & -i  \cr 0 & 0 & 0 & 0 \cr 0 & i & 0 & 0  }\,,\;\;
\\
\\
 \mu^-_6 =   \sqrt { n-1 \over 2n } \; \pmatrix { 0 & 0 & 0 & 0 \cr 0  & 0 & 0 & 0 \cr 0 & 0 & 0 & 1 \cr 0 & 0 & 1 & 0  }\,,\;\;
 \mu^-_7 =    \sqrt { n-1 \over 2n } \; \pmatrix { 0 & 0  & 0 & 0 \cr 0  & 0 & 0 & 0 \cr 0 & 0 & 0 & -i \cr 0 & 0 & i & 0  }\,.
\EQ
The top left corner of the $\mu^+_i$ matrices are proportional to the antielectron matrices (3.8)
\BQ
 \mu^+_4 =   { \sqrt {n+1 \over 2n}} \; \pmatrix { 0 & 0 & i & 0 \cr 0 & 0 & 0 & 0 \cr -i  & 0& 0 & 0 \cr 0 & 0 & 0 & 0  }\,,\;\;
 \mu^+_5 =   { \sqrt {n+1\over 2n }} \; \pmatrix { 0 & 0 & 1 & 0\cr 0 & 0 & 0 & 0 \cr 1 & 0 & 0 & 0 \cr 0 & 0 & 0 & 0  }\,,\;\;
\\
\\
 \mu^+_6 =   { \sqrt {n+1 \over 2n }} \; \pmatrix { 0 & -i  & 0 & 0 \cr i & 0 & 0 & 0 \cr 0 & 0 & 0 & 0 \cr 0 & 0 & 0 & 0  }\,,\;\;
 \mu^+_7 =   { \sqrt {n+1 \over 2n }} \; \pmatrix { 0 & -1 & 0 & 0 \cr -1 & 0 & 0 & 0 \cr 0 & 0 & 0 & 0\cr 0 & 0 & 0 & 0  }\,.
\EQ
If we sum over a complete family, we find that the matrices
associated to $H$ and $K$ have the same trace-metric:
\BE
   Tr (\mu^+_i \mu^-_j) = 0\;,\;\;Tr (\mu^+_i \mu^+_j) = Tr (\mu^-_i \mu^-_j) = (n + 1)\; \delta_{ij}\,, \qquad i,j = 4,5,6,7\,.
\EE
This identity, like (4.8), is a consequence of the vanishing of the sum of the
charges of all the right singlets (electron + n up quarks  + n down quarks) of a family.
It shows that one can compute the renormalization of the scalar 
propagator either using the $\Phi_L,\;\Phi_R$ fields (5.4) or using the $H,\;K$ fields (5.8), and 
obtain the same result (5.9).

\section {The scalar potential}

From now on, we restrict our attention to the case $n=3$ colors.
Consider the structure of the scalar potential induced
by a Fermion loop. The cubic terms vanish, since each interaction with a scalar
changes the chirality of the Fermions. The term of degree 6 or higher converge. 
So we only have to compute the quartic potential.
If we work in terms of the $\Phi$ fields, the counterterm is proportional to
\BQ
\Phi^i_L\;\Phi^j_R\;\Phi^k_L\;\Phi^l_R\;\;Tr_L (\lambda_i\;\lambda_j\;\lambda_k\;\lambda_l\;)\;.
\EQ
Generalizing (5.6), we find that many unfriendly terms, for example 
$\Phi^4_L\;\Phi^4_R\;\Phi^4_L\;\Phi^5_R$, are eliminated by the balance
between the leptons and the quarks. 
If we introduce the metric:
\BE
 \GT_{ab} =  \pmatrix { 2 \delta_{ab} - \alpha\;\gHH_{ab}&  - \gHH_{ab} \cr - \gHH_{ab} & 2 \delta_{ab} - \alpha\;\gHH_{ab}}\;,
\EE
and the products
\BQ
 	Z^a_L = (f^a_{ij} - i\;d^a_{ij}) \;\Phi^i_L\,\Phi^j_R\;,\\
 	Z^a_R = (f^a_{ij} + i\;d^a_{ij}) \;\Phi^i_L\,\Phi^j_R\;.
\EQ
where the $d^a_{ij}$ and the $f^a_{ij}$ are defined in (3.4), 
the direct calculation of the Fermion loop induces, up to an infinite
multiplicative constant, a scalar potential of the form
\BQ
 V =  {1 \over 8 }\; \pmatrix {Z^a_L & Z^a_R} \; \GT_{ab}\; \pmatrix {Z^a_L \cr Z^a_R}\;.
\EQ
The $\alpha$ parameter is arbitrary in (6.4), because we have the identity
 \BE
   Z^a_L \; \gHH_{ab}\;Z^b_L = 0\;.
\EE
If we work with the $\mu^{\pm}$ Hermitian matrices (5.12-14), we find,
generalizing (5.15), that the symmetrized quartic traces are identical
\BQ
Tr ((\mu^+_i \mu^+_j + \mu^+_j \mu^+_i)(\mu^+_k \mu^+_l + \mu^+_l \mu^+_k))
\\
=  Tr ((\mu^-_i \mu^-_j + \mu^-_j \mu^-_i) (\mu^-_k \mu^-_l + \mu^-_l \mu^-_k))
\\
= 32/9\;(\delta_{ij} \delta_{kl} +  \delta_{ik} \delta_{jl} +  \delta_{il} \delta_{jk}) \;.
\EQ
Finally the mixed traces satisfy, for all $i,j,k,l$:
\BQ
   Tr (\mu^+_i \mu^-_j) = 
   Tr (\mu^+_i \mu^-_j \mu^-_k \mu^-_l) =
   Tr (\mu^+_i \mu^+_j \mu^+_k \mu^-_l) = 
   Tr (\mu^+_i \mu^-_j \mu^+_k \mu^-_l) = 0 \,,
\\
\\
 Tr (\mu^+_i \mu^+_i \mu^-_i \mu^-_i) = 0 \;,\qquad \hbox{for any i, no sum implied.}
\EQ
If we now introduce the symmetric product
\BQ
 	(X * X)^a = d^a_{ij} \;X^i\,X^j\;;\;\;\;X = H,\; K\;,
\EQ
the direct calculation of the Fermion loop induces, up to the same infinite
multiplicative constant, a scalar potential of the form
\BE
 V =  {1 \over 16 }\;\pmatrix {(H * H)^a &  (K * K)^a}  \pmatrix { \GT_{ab}} \pmatrix {(H * H)^b \cr  (K * K)^b}\;. 
\EE
The $\alpha$  parameter (6.2) is arbitrary in (6.9), because in any Lie superalgebra,
we have the identity
 \BE
   \gHH_{ab} (X * X)^a \;(X * X)^b = 0
\EE
Notice that for the same reason, the mixed $H\;K$ term vanishes 
if and only if $H$ and $K$ are parallel. 
This identity insures that the centralizer of the true vacuum is nontrivial,
or in other words, that the photon remains massless.

If we change variable again and define
\BQ
  u_1 = (H_4 + i H_5)/\sqrt 2\;,
\\
  u_2 = (H_6 + i H_7)/\sqrt 2\;,
\\
  v_1 = (K_4 + i K_5)/\sqrt 2\;,
\\
  v_2 = (K_6 + i K_7)/\sqrt 2\;,
\EQ
and substitute in (5.9) and (6.9), we recover up to a scale factor $\sigma$ the  propagator and potential 
of Fayet \cite{F74,F75}

\BQ
\LAG_s + V =  \partial_{\mu} \overline{u}\; \partial_{\mu} u +
      \partial_{\mu} \overline{v}\; \partial_{\mu} v\; +
\\
+  {\sigma} ((\overline{u} u)^2 +  (\overline{v} v)^2 + 
    (\overline{u}_1 \overline{v}_2 - \overline{u}_2 \overline{v}_1)(u_1 v_2 - u_2 v_1))\;.
\EQ
In these variable, it is immediately clear that, assuming a symmetry breaking negative mass 
term for the scalars, the potential is minimal when $u$ and $v$ are
parallel, and therefore the photon remains massless. 
The scalar potential (6.9) is invariant under the infinitesimal transformation
\BQ
\Delta\;H_i = (H^k\;\delta_{kl}\;K^l)\;\gHH_{ij}\;K^j - (H^k\;\gHH_{kl}\;K^l)\;\delta_{ij}\;K^j\;\;,
\\
\Delta\;K_i = (H^k\;\delta_{kl}\;K^l)\;\gHH_{ij}\;H^j - (H^k\;\gHH_{kl}\;K^l)\;\delta_{ij}\;H^j\;\;,
\EQ
or, if we use the variables $u$ and $v$:
\BQ
\Delta\;u = (u . {\overline v})\; v\;\;,\;\;\;
\Delta\;v = - (v . {\overline u})\; u\;\;,
\\
\Delta\;{\overline v} = (u . {\overline v})\; {\overline u}\;\;,\;\;\;
\Delta\;{\overline u} = - (v . {\overline u})\; {\overline v}\;\;,
\EQ
whereas the term $(u.v)({\overline u} . {\overline v})$ which could give a mass to the photon and is 
absent from our Lagrangian, is not invariant under $\Delta$.

\section {Scalar Ward identities}

The study of the counterterms induced by the Fermion loop has given in the previous section
the structure of the propagator and of the potential, but not their scale. If we write our Lagrangian
as 
\BQ
	\LAG = -{1 \over 4} (F^a_{\mu\nu})^2
	- {1 \over 2} ((D_{\mu}H)^2 + (D_{\mu}K)^2) + \lambda\;{g^2}\;V 
	+ {\overline \Psi}\; {g \over 2}\;(\gamma^{\mu}\;A_{\mu} + \xi \Phi\;)\; \Psi\;,	
\EQ
The factor $2$ in $g/2$ is needed because our $SU(2)$ matrices (3.1) are normalized
as usual in mathematics, not as the standard spin $1/2$ isospin matrices of physics.
The weak angle and the coefficients $\lambda$ and $\xi$ are not yet known. 
We will now fix them by requiring that they
do not depend on the number of lepton families. 

The only contribution of the other families to the coupling of the vectors and scalars to the
Fermion of the first family is through the renormalization of the wave function of the
vector and the scalar. The counterterms are proportional to
\BQ
     Z(A_a^2) = {2 \over 3}\;{g^2 \over 4} \;Tr_f\;\lambda_a^2\;\;,\;\;\;
     Z(A_8^2) = tg^2\theta\;{2 \over 3} {g^2 \over 4} \; Tr_f\;\lambda_a^2\;\;,\;\;\;
     Z(H^2) = \xi^2\;{g^2 \over 4} \; Tr_f\;\mu_i^2\;\;.
\EQ
The factor $2/3$ in the vectors comes from the kinematic contribution of the Dirac matrices.
We count $2/3$, not the usual $4/3$ of a massive Fermion, since here the left and right spinors
contribute separately to the trace term.
Over a full family, the $SU(2)$ even matrices (3.1,4.1) are normalized to 
$Tr_f(\lambda_a\;\lambda_b)=8\delta_{ab}\;$. The $U(1)$ matrix is
normalized to $Tr_f(\lambda_8\;\lambda_8)=40/3$, 
and the odd matrices to $Tr_f(\mu^{\pm}_i\;\mu^{\pm}_j)=4\;\delta_{ij}$.
To integrate these 3 counterterms in the renormalization of the coupling constant $g$,
and insure that $tg\theta$ and $\xi$ are not renormalized,
we fix 
\BQ
tg^2\theta = {3 \over 5}\;\;,\;\;\;sin^2\theta = {3 \over 8}\;\;,\;\;\;\xi = {2 \over \sqrt{3}}\;\;.
\EQ
The weak angle coincides with the value derived from the $SU(5)$ grand unified theory \cite{GG74}.
We then look at the renormalization of the scalar potential. The standard Lie
algebra Ward identities ensure that the contribution of the Fermions to the self coupling
of the vectors, the $g^2 \;A^4$ term, can be reabsorbed in the renormalization
of $g^2$. A similar condition holds for the scalar potential $\lambda\;g^2\;H^4$.
\BQ
Z(V) = {1 \over 3}\,{(g\xi)^4 \over \lambda g^2} 
\EQ
The contribution of the Fermions is reabsorbed in the renormalization of $g^2$
and therefore $\lambda$ is not renormalized if we choose
\BQ
Z(V) = Z(H^2) ==>
\lambda = {\xi^2 \over 3} = {4 \over 9}
\EQ 

However, we note that the Bosonic counterterms, induced by the vector and scalar 1-loops
perturb all the scalar couplings, except the very structure of the scalar potential
which guarantees the survival of an unbroken $U(1)$ and the existence of the massless photon.
Furthermore the
strong interactions gluon vector fields interact with the quarks, but not with the leptons.
The photon-quark coupling is protected by a Ward identity and is not renormalized by
the gluon-quark interaction, but the scalar-quark vertex is not protected. Hence, the
balance between quarks and leptons, which is central to our analysis is broken by these
counterterms. This problem can either be interpreted  as a fatal flaw of the $SU(2/1)$ model
or as sign that the quantum theory is still incomplete.

\section {The mass of the Higgs}

Since we have now fixed the scale of the potential we can derive the
mass of the physical scalars, the two neutral fields $H^0_{1 or 2}$ and 
the charged field $H^{\pm}$. Suppressing 
the Fermions, the Lagrangian reads:
\BQ
	\LAG = -{1 \over 4} (F^a_{\mu\nu})^2
	- {1 \over 2} ((D_{\mu}H)^2 + (D_{\mu}K)^2) + {\lambda}\;V
\EQ
where $V$ is given in (6.9). If we develop the potential and
show explicitly the occurrences of the component fields $H_4$.
We find:
\BQ
 {\lambda}\;V =  {\lambda \over 4}\; (H^4_4 +
 H^2_4\;(2\;(H^2_5 + H^2_6 + H^2_7) + K^2_6 + K^2_7) + 
\\
+ 2\;H_4 H_6\; (K_4 K_6 + K_5 K_7) +  2\;H_4 H_7\; (K_4 K_7 + K_5 K_6)) + ...
\EQ
Three points should be noticed. The $\lambda$ parameter is scaled as usual to give $\lambda/4\;H^4_4$.
There are no terms $H^2_4\;K^2_4$ or  $H^2_4\;K^2_5$, finally the coefficients of $H_4$ are invariant
under a $\lambda_8$ rotation of the $K$ fields, which rotates $K_4$ into $K_5$ and $K_6$ into $K_7$.

We now add by hand in the Lagrangian a quadratic symmetry breaking 'imaginary mass' term
\BQ
- {\lambda \over 2} \; v^2\; (sin^2\beta\;H^2 + cos^2\beta\;K^2)\;.
\EQ
We assume that the vacuum expectation value of $H$ is in the direction $H_6$. As discussed
in section 3, this choice just fixes the name of the particles. To minimize
the mixed $H^2\;K^2$ potential, the $K$ vacuum expectation value must be in
the plane $(K_6,K_7)$. The vacuum correctly preserves the photon (3.5, 4.7). We then rotate
$<K>$ to $<K_6>$. This 'phase' invariance corresponds to the fact that one of the 4
neutral states in the hyper-plane $(H_6,H_7,K_6,K_7)$, usually called the '$A_0$'
pseudo-scalar, is absent from the quartic potential and its mass can be fixed arbitrarily
in the symmetry-breaking term. We minimize the potential and find:
\BQ
	<H_6> = v\;sin \beta \;\;,\;\;\;<K_6> = v\;cos \beta \;\;,\;\;\;M_W = v\;g/2\;,
\EQ
where $\beta$ is defined modulo $\pi / 2$. We then expand the potential around the vacuum to
second order in $v$. We find 4 massless states which produce the $A_0$ and the 
longitudinal components of the massive vector fields $Z$ and $W^{\pm}$ 
\BQ
    H_7\;,\;\;K_7\;,\;\;(sin\beta\;H_4 + cos \beta K_4) \pm i (sin\beta\;H_5 + cos \beta K_5)\;,
\EQ
two charged physical scalars
\BQ
   H^{\pm} = (cos\beta\;H_4 - sin \beta K_4) \pm i (cos\beta\;H_5 - sin \beta K_5)\;,
\\
M_{H^{\pm}} = {\sqrt {2\;\lambda}}\;M_W = 75.8\; GeV
\EQ
and two neutral scalars in the $(H_6,K_6)$ plane, usually called $H^0_1$ and $H^0_2$, with masses
\BQ
 M_{H^0_1} = {2\;\sqrt {2\;\lambda}} \;cos \beta\;M_W = ({\sqrt 2} cos\beta)\;107.2\;GeV\;,
\\
 M_{H^0_2} = {2\;\sqrt {2\;\lambda}} \;sin \beta\;M_W = ({\sqrt 2} sin\beta)\;107.2\;GeV\;.
\EQ
In the symmetric case $\beta = \pi / 4$ the 2 neutral scalar are at $107.2\;GeV$.

\section{The $SU(2/1)$ hyper-curvature}

Since the scalar fields (5.3) play the role of a gauge field associated to
the odd generators of the superalgebra, it seems natural to introduce
them as part of the covariant differential and curvature 2-form.
Such a construction was proposed in \cite{TM06}, where we found that
the covariant differential is associative if and only if the
Fermions sit in a representation of the superalgebra graded by
the chirality. As shown in sections 2 and 3, this condition is met by
the observed leptons and quarks. But we found in section 3 the additional
complication that the quark representation is non Hermitian, so
the scalar field is complex, so effectively, the number of odd generators
in the quantum field theory is twice the number of odd generators of the
$SU(2/1)$ superalgebra. 

We are going to show that the construction of the curvature can be generalized
and yields a structure we would like to call a H-curvature,  by reference to 
Fayet and Iliopoulos \cite{F74,F75}, who were the first to consider, 
in the context of Hyper-Symmetry, a doubling 
of the Higgs field introduced by Weinberg \cite{Wei67} 
and to Sternberg and Wolf \cite{SW78} who introduced the Hermitian algebras.

Following \cite{TM06}, we define the Dirac 1-forms
\BQ
\GX = dx^{\mu}\;\sigma_{\mu}\ + dx^{\mu}\;\overline{\sigma}_{\mu}\,.
\EQ
where $\sigma$ maps the right Fermions on the left Fermions and vice versa.
We then define the chiral Yang-Mills 1-form $\AT$ and covariant differential $\DT$
\BQ
\AT = A + \Phi_L + \Phi_R = dx^{\mu}\;A^a_{\mu}\;\lambda_a + \GX\;\Phi^i_L\lambda_{iL} 
	+ \GX\;\Phi^i_R\lambda_{iR}\qquad ,
\DT = d + \AT\;.
\EQ
Notice that $\AT$ is an exterior form of homogeneous degree 1, in contradistinction
with the Ne'eman Quillen connection \cite{TMN82,Q85} sometimes discussed
in the $SU(2/1)$ literature (\cite{NS90,NSF05} and references therein).
The curvature is defined as the square of the covariant differential
\BQ
\FT = \DT\;\DT = d\AT + \AT\;\AT\;.
\EQ
Since the $dx^{\mu}$ anticommute, the classical term 
\BQ
A\;A = dx^{\mu}dx^{\nu}\;A^a_{\mu}\;A^b_{\nu}\;\lambda_a\lambda_b
\EQ 
is antisymmetric in $ab$ and is internal provided the even matrices $\lambda_a$
close by commutation
\BQ
\lambda_a \lambda_b - \lambda_b \lambda_a = i\;f^c_{ab} \;\lambda_c\;.
\EQ
Similarly, the scalar vector term in $\GX\;D\Phi$ generate the covariant differential
of the scalars provided the even odd sector close by commutation
\BQ
\lambda_a \lambda_i - \lambda_i \lambda_a = i\;f^j_{ai} \;\lambda_j\;.
\EQ
However, the structure of the scalar-scalar term is more involved. Since
$\Phi_L$ and $\Phi_R$ are independent, the $\lambda$ matrices are neither symmetrized
nor skew symmetrized and we get
\BQ
{1 + \chi \over 2} \;\Phi^i_L\Phi^j_R\; \lambda_i \lambda_j + 
{1 - \chi \over 2} \;\Phi^i_R\Phi^j_L\; \lambda_i \lambda_j\;.
\EQ
We now decompose the matrix product into the symmetric and skew symmetric parts
\BQ
 \lambda_i \lambda_j = {1 \over 2} (\{\lambda_i, \lambda_j\} + [\lambda_i, \lambda_j])
\EQ
and we find that the scalar-scalar term can be expressed as a linear combination of even
generators provided the anticommutator of the odd matrices close on the
even matrices and the  commutators of the odd matrices close on $\chi$ times the even matrices,
defining the skew-symmetric structure constants $f^a_{ij}$:
\BQ
 \lambda_i\, \lambda_j + \lambda_j\, \lambda_i = d^a_{ij}\,\lambda_a\;,
\\
 \lambda_i\, \lambda_j - \lambda_j\, \lambda_i = i\;\chi\;f^a_{ij}\,\lambda_a\;.
\EQ
Therefore the unsymmetrized product of the odd matrices satisfy
\BQ
 \lambda_i\, \lambda_j  = {1 \over 2} (d^a_{ij} + i\;\chi\;f^a_{ij})\lambda_a\;.
\EQ
By inspection (3.4), all these conditions are satisfied by the lepton representation (3.1-3).
In fact, the construction works with the fundamental representation
of any $SU(m/n)$ superalgebra, because all the generators coincide
with the generators of $SU(m+n)$ except for the supertraceless $U(1)$
but $\chi\;U(1)$ is traceless and coincides with the $U(1)$ of
$SU(m+n)$.

The structure (9.9) was introduced by Sternberg and Wolf \cite{SW78}, who call it
 a Hermitian algebra and discussed by Ne'eman and Sternberg in several subsequent 
papers \cite{NS90,NSF05}, but its exact relevance to quantum field theory 
remained unclear. The unusual presence of the chirality operator $\chi$ 
on the right hand side of the
commutator (9.9) was not explained, because they did not consider the possibility
of splitting the odd generators and doubling the number of Higgs fields. But in our
modified framework, if we compare (9.7) and (9.10) and use $\chi^2 = 1$, 
we see that the $\chi$ operator present in (9.9) disappears from the
curvature 2-form which is now correctly valued on the even matrices and can be written as
\BQ
\FT_L = \FHH^a_L\;\lambda_a  + \GX\;D\Phi^i_L\;\lambda_{iL} + \GX\;D\Phi^i_R\;\lambda_{iR}\;,
\\
\FHH^a_L = dA^a + {i \over 2} f^a_{bc}\; A^bA^c +  {i \over 2} \;\GX\GX \; \Phi^i_L\Phi^j_R \;(f^a_{ij} - i\;d^a_{ij}) \;.
\EQ
The Bianchi identity $\DT\;\FT = (\DT\DT)\;\DT - \DT\;(\DT\DT) = 0$ follows from the associativity of the
matrix product and implies the Hermitian Jacobi identity of \cite{SW78}.
We can also construct a second solution $\FHH_R$ by switching the sign of the $d$ constants
changing the chirality:
\BQ
\FHH^a_R = dA^a + {i \over 2} f^a_{bc}\; A^bA^c +  {i \over 2} \;\GX\GX \; \Phi^i_L\Phi^j_R \;(f^a_{ij} +i\;d^a_{ij})\;.
\EQ
We recover in (9.11-12) the object $Z_{L/R}$ introduced in (6.3)
\BQ
\FHH_{L/R} = F + {i \over 2} \;\GX\GX \;\;Z_{L/R}
\EQ

\section{The electro-weak angle}

From the definition of the classical curvature 2-form given in the previous section,
we can construct the Lagrangian in the usual way, using the Hodge adjunction $^*$
and the quadratic form $\GT_{ab}$ (6.2) which reproduces the scalar potential induced 
by the Fermion loops. 
The Lagrangian can be written as:
\BE
 {\LAG}_F 
  =  {1 \over 8\;g^2} Tr \;\pmatrix {^*\FHH^a_L & ^*\FHH^a_R}   \GT_{ab}  \pmatrix {\FHH^b_L \cr \FHH^b_R}
\EE

Since the new bilinear scalar term in the H-curvature (9.13) coincides with (6.3), the scalar potential is proportional to (6.4) which can be rewritten as (6.9).
It is curious to see that  the combined effect of the leptons and quarks in (6.9) is mimicked 
in the classical construction (10.1) by combined effect of the 2 chiral ways (9.11-12) of extending
the superalgebra $SU(2/1)$ curvature of \cite{TM06} to the H-algebra.

The interesting point is that, using (6.2) and (3.6), the effective metric for the Yang-Mill 
curvature $F^a$ comes out as
\BE
2\; g_{ab} - (1 + \alpha)\;\gHH_{ab} = \hbox{diag} (1-\alpha,1 - \alpha,1 - \alpha,3+\alpha)\,,\qquad a,b = 1,2,3,8
\EE
and we learn the choice of the $\alpha$ parameter in the definition of the $\GT$ hyper-metric corresponds
to the choice of the electro-weak angle in the classical Lagrangian according to:
\BQ
tg^2\theta = (1 - \alpha) / (3 + \alpha)
<=>
\alpha = 1 - 4\;sin^2\theta\;.
\EQ
Substituting (7.3), we find $\alpha = - 1/2$.

\section{Discussion}

The $SU(2/1)$ model appeared in 1979 as very compelling, but so far no setting
had been found that could explain how to extend the Yang-Mills quantum field
theory from gauging an internal Lie algebra symmetry to the case of
a superalgebra. We believe that the present paper
represents a step in this direction.

The original observation \cite{N79,F79} is that,
provided we grade the Fermions by their chirality,
the smallest Lie superalgebra 
$SU(2/1)$ exactly corresponds to the standard model.
The leptons and the quarks fit the 2 smallest irreducible 
representations (3.1-2) and (4.1-4) of the superalgebra $SU(2/1)$.
The $U(1)$ hyper-charge, which is supertraceless, 
coincides with the original choice of Weinberg \cite{Wei67}
who wanted to avoid a current coupled to the lepton number, and
explains the non existence of any massless charged particle.
The difficulty is to understand the role of the odd generators.

Our new idea is to take the $SU(2/1)$ matrices at face value,
despite the fact that they are not Hermitian (4.3), and to construct
the whole Lagrangian by studying the counterterms
induced by the Fermions. The caveat is to remember
the triangle anomaly (4.5). It is avoided
if and only if we always consider as our building block 
a whole family, constituted of one lepton triplet and
three quark quadruplets, for example the electron, the left
neutrino and 3 colored copies of the up and down quarks,
counting independently the left and right chiral states.
If we associate a single real scalar to each odd generator
the induced scalar propagator vanishes (4.8). But if 
we split the odd matrices into their chiral parts (5.2),
and consider 2 sets of scalars $\Phi_L$ and $\Phi_R$  (5.3)
the leptons and the quarks conspire
to induce a the scalar propagator (5.7) proportional
to the odd part of the skew-symmetric Killing metric of the superalgebra (3.6).
It is then possible to change variables (5.8) from $\Phi_L,\;\Phi_R$ to $H,\;K$
and recover two Higgs scalar doublets,
$H$ coupled to the electron (5.12) and down quark (5.13), 
and $K$ coupled to the up quark (5.14).
If we compute in the same way the quartic scalar counterterms
we recover (6.12) the potential first considered by Fayet \cite{F74},
a building stone of what later became the Minimal Standard Supersymmetric Model (MSSM).
This is rather surprising, because we have not introduced 
any Wess-Zumino supersymmetry in our construction, but may be
attributed to the near unicity of the $SU(2)U(1)$ quartic invariants.

Some interesting results follow from the analysis of the
quantum stability of the symmetry breaking pattern of the theory.
In the Fayet potential, the crucial element is the form of the
mixed term which insures that, in the vacuum, the 2 fields
$H$ and $K$ are parallel, and therefore leave an unbroken $U(1)$
symmetry, gauged by a massless photon. In our model, this
condition is a consequence of the super-Jacobi identity
(6.10) and is stable against the Fermionic 1-loop
quantum corrections, since this is the way we derived
the scalar potential. 
Furthermore, if fix (7.3) the relative strength of the 2 vectors
coupling to $tg^2\theta=5/3$ (as in $SU(5)$ grand
unified theory \cite{GG74}), the strength of the scalar
to $2/{\sqrt 3}$, and the quartic self coupling to
$\lambda=4/9$, these coefficients
are stable against
the quantum correction induced by the Fermion loops for
any number of families, thanks to the Ward identities (7.5).
These identities are typical of a Yang-Mills minimal
coupling and states that scalar Fermion coupling and the scalar
quartic self coupling are indeed proportional to $g$ and $g^2$
where $g$ is the Yang-Mills $SU(2)$ coupling constant.
This fixes (8.7) the mass of the neutral Higgs at
 $M^2_{H^0_1} + M^2_{H^0_2} = 2\;(4/3\;M_W)^2 = 2\;(107.2\;GeV)^2$.
The mass of the charged Higgs corresponds to the mixed $H$ $K$ term
in the potential and gives (8.6):
$M_{H^{\pm}} = 2 {\sqrt 2}/3\;M_W = 75.8\; GeV$, a value a few $GeV$ too low
relative to the minimum given by the particle data group \cite{Yao06}.
Finally the mass of the '$A_0$' pseudo-scalar is not constrained.

On the other hand, we clearly have a difficulty with the mass of
the quarks. If we assume that the 2 Higgs have the same mass
and that the $SU(2/1)$ minimal coupling (5.3) directly gives the
mass of the heaviest quark, we predict $M_{top} = M_{H^0} = 4/3\;M_W = 107.2\; GeV$
which falls way under the current experimental value
$M_{top} = 174.3 \pm 1.8 \; GeV$, deduced from direct observations \cite{Yao06}.
But a worse problem is that the $charm$ and $up$ quarks, in the
two other families should have the same mass, when actually they
are very light ($1.25 \; GeV$ and $2 \; MeV$). 
This problem is possibly solved by the $SU(2/1)$ indecomposable representations
which have been shown by Haussling and Scheck to correctly
reproduce the Masakawa quark mixing phenomenology \cite{HS94}.
Assuming that the masses get redistributed, we are not so far from
the relation $M^2_{top} + M^2_{charm} + M^2_{up} = 3\;M^2_H$.
This question needs further studies.

A classical geometric interpretation of the theory is also possible.
When expressed in terms of the $\Phi_L$ and $\Phi_R$ fields, we found that
the potential involves both the symmetric $d^a_{ij}$ structure constants of
the superalgebra and the skew symmetric $f^a_{ij}$
structure constants of $SU(3)$, times the chirality operator (3.4).
This gives the idea of revisiting
the Hermitian algebras introduced by Sternberg and Wolf \cite{SW78}.
Our new result is to show that this structure exactly fits
the chiral decomposition (5.3) of the $\Phi$ fields, allowing us
to construct a classical curvature 2-forms (9.12), where the unexpected
apparition of the chirality operator $\chi$ in the definition
of the Hermitian Lie bracket (9.9) exactly compensates the signs
implied when adapting our chiral connection 1-form $\AT$ \cite{TM06}
to the doubling of the $\Phi$ fields. There is also a probable relation 
with non commutative differential
geometry, as discussed in \cite{TM06}.

We are now in a position to give a classical geometrical construction of 
the vector/scalar part of our
Lagrangian as the square of the curvature of the $SU(2/1)$ Hermitian 
algebra. In the earlier work of Ne'eman and others, the central
problem was the choice of the quadratic form needed to compute
the square of the curvature. The natural choice was the supertrace
metric (3.6), but this yields a negative propagator for the $U(1)$ field
and breaks unitarity. The phenomenological choice seemed to be the
trace of the lepton representation (3.1) which yields $sin^2\theta = .25$.
In 1979, this value was acceptable, but is now too high. Also
the trace metric is not a good invariant for a superalgebra, and this
choice was often criticized. Our new result is to deduce the relevant
quadratic form from the renormalization theory. We found the very elegant
hyper metric $\GT$ (6.2) which is a combination of the trace and the
supertrace metric and, thanks to the super-Jacobi identity, contains
a free parameter $\alpha$. The 2 rows and columns of $\GT$ are used to
combine the 2 $CP$ conjugated versions of the H-curvature (10.1), and this
combination restores unitarity, because in the combined Lagrangian we can
switch between the native $SU(2/1)$ geometric but non-Hermitian $\Phi_L$ and  $\Phi_R$ (6.4)
variables and the good Hermitian quantum field theory scalars $H$ and $K$ (6.9).
When we inject this metric in our Lagrangian (10.1),
we find that the freedom in the choice of $\alpha$ reflects a freedom
in the choice of the electro-weak angle  $\alpha = 1 - 4\;sin^2\theta$ (10.3)
and that the simplest choice $\alpha = 0$
corresponds to the trace-metric choice, $\alpha = -1/2$ corresponds to $sin^2\theta=3/8$.

Let us now recall the distinction between this model and the minimal
supersymmetric standard model (MSSM). The geometrical space behind
Yang-Mills theory is the principal fiber bundle. The base is space-time, 
the fiber is the gauge group. In both cases, a superalgebra is introduced
but not in the same way. In the MSSM, the Poincare superalgebra is
introduced in the base space, acts on Bose/Fermi supermultiplets
and naturally leads to supergravity.
In our approach, the superalgebra is introduced in the fiber,
by embedding the Glashow-Weinberg-Salam $SU(2)U(1)$ Lie algebra in $SU(2/1)$,
it acts of left/right Fermionic chiral supermultiplets and
groups the Higgs scalars with the Yang-Mills vectors. This
construction is not incompatible with supergravity at much higher energy
or with superstrings but does not lead to these theories. 
In both models, (5.11) and \cite{F74}, we
are led to introduce 2 sets of scalar doublets, $H$ coupled to the
electron and down-quark right singlets, and $K$ to the up-quark singlet.
The quartic scalar potential has the same structure (6.9-12) and implies
the survival of an unbroken $U(1)$ and a massless photon. So the scalar
spectrum is similar, 3 states become the longitudinal components of the $W$ and $Z$ vector fields, 5 survive.
In both models the scale of the potential is fixed and the values are
in the same range. But of course the main difference is that
the MSSM predicts for every known particle the existence of supersymmetric partner, 
like the winos, gluinos, squarks and sleptons, whereas $SU(2/1)$ fits
exactly the known particles but does not predict any new ones, except the Higgs
fields. The existence or non existence of these particles below $1\; TeV$ will
allow us to choose between the 2 kinds of supersymmetry.

The $SU(2/1)$ theory is testable, but incomplete in several directions. 
At the experimental level, the model predicts
2 complex doublets of Higgs fields, rather than 1 as
in the original Weinberg paper, and constrains their masses
within the range of the new CERN collider which opens this year.
At the phenomenological level, the mixing of the quark families
is not understood in details, but may be linked to the
indecomposable representations of $SU(2/1)$.
At the theoretical level, the balance
between the quarks and leptons is broken by the 
quantum counterterms induced by the gluons,
and all the scalar couplings, except those insuring the
existence of a massless photon, are perturbed by the
Boson loops. We believe that this problem could indicate the
incompleteness of the current quantization procedure and
will discuss it elsewhere.

But even if these difficulties  cannot be solved
immediately, the 1-loop stability of the structure of the scalar
potential, ensuring a massless photon, and the resulting prediction
of the number and masses of the Higgs fields seem worth
reporting as a first step towards the construction of
$SU(2/1)$ Quantum Astheno Dynamics.

\bigskip

\acknowledgments

In many ways, this paper is a concretization of 
our long lasting collaboration with Yuval Ne'eman on SU(2/1).
It is also a pleasure to thank Pierre Fayet and Victor Kac
for many discussions, Florian Scheck and Shlomo Sternberg
for communications, David Lipman for always prompting new ideas and
Danielle Thierry-Mieg for important suggestions.


\begin{thebibliography}{999}
\bibitem[AL73]{AL73} E.S. Abers and B.W. Lee, ``Gauge Theories",
{\em Physics Reports}, {\bf 9C} no. 1 1973
\bibitem[BIM72]{BIM72} C.Bouchiat, J.Iliopoulos and P.Meyer,
An anomaly free version of Weinberg's model
{\em Phys. Let.} {\bf A38}, 519 (1972)
\bibitem[CL90]{CL90} A. Connes and J. Lott, ``Particle models and
noncommutative geometry", {\em Nucl. Phy.} {\bf 18B} (Proc.
Suppl.) 29 (1990)
\bibitem[DJ79]{DJ79} P.H. Dondi and P.D. Jarvis,
``A supersymmetric Weinberg-Salam model", {\em Physics Letters}
{\bf 84B} Erratum {\bf 87B} 403 (1979)
\bibitem[F79]{F79} D.B. Fairlie, ``Higgs Fields and the Determination of the
Weinberg Angle", {\it Phys. Lett.} {\bf B82} 97 (1979)
\bibitem[F74]{F74} P.Fayet,  {\em Nucl. Phy.} {\bf B78}, 14 (1974)
\bibitem[F75]{F75} P.Fayet,  {\em Nucl. Phy. Lett.} {\bf B90}, 104 (1975)
\bibitem[GG74]{GG74} H.Georgi and S.Glashow,
{\em Phys. Rev. Lett.} {\bf 32},  (1974)
\bibitem[GIM70]{GIM70} S.Glashow, J.Iliopoulos and L.Maiani,  
{\em Phys. Rev.} {\bf D2}, 1285-1292 (1970)
\bibitem[BIM72]{BIM72} C.Bouchiat, J.Iliopoulos and Ph.Meyer 
{\em Phys. Lett.} {\bf B38}, 519-523 (1972)
\bibitem[HS94]{HS94} R.Haussling and F.Scheck, ``Quark mass matrices and 
generation mixing in the standard model with non-commutative geometry", 
{\em Phys. Lett.} {\bf B336}, 477-486 (1994)
\bibitem[N79]{N79} Y. Ne'eman, ``Irreducible Gauge Theory
of a Consolidated Salam-Weinberg Model", {\em Phys. Lett.} {\bf
B81}, 190 (1979)
\bibitem[NS90]{NS90}Y. Ne'eman, S. Sternberg,
``Superconnections and internal supersymmetry dynamics", {\em
Proc. Nat. Acad. Sci. USA} {\bf 87} 7875 (1990)
\bibitem[NSF05]{NSF05} Y. Ne'eman, S.Sternberg and D.Fairlie ``Superconnections 
electroweak su(2/1) and extensions, and the mass of the Higgs", {\em Physics Reports} {\bf
406}, 303-377 (2005)
\bibitem[NT80]{NT80} Y. Ne'eman and J. Thierry-Mieg,
``Geometrical Gauge Theory of Ghost and Goldstone Fields and of
Ghost Symmetries", {\it Proc. Nat. Acad. Sci., USA} {\bf 77}
720-723 (1980)
\bibitem[Q85]{Q85}D. Quillen, ``Superconnections and the Chern
character", {\em Topology} {\bf 24} 89 (1985)
\bibitem[Sa68]{Sa68} A. Salam, in {\it Elementary Particle
Theory}, N. Svartholm, ed. Almquist Verlag A.B., Stockholm (1968)
\bibitem[S92]{S92} F.Scheck, ``Anomalies, Weinberg angle and a 
non-commutative geometric description of the standard model",
{\em Phys. Lett.} {\bf B284}, 303-308 (1992)
\bibitem[SNR77]{SNR77} M. Scheunert, W. Nahm \& V. Rittenberg,
``Irreducible Representations of the osp(2/1) and sp1(2/1) graded
Lie algebras", {\it J. Math. Phys.} {\bf 18} 156-167 (1977)
\bibitem[SW78]{SW78} S. Sternberg, J. Wolf, ``Hermitian Lie
algebras and metaplectic representations", {\em Trans. Amer. Math.
Soc.} {\bf 231} 1 (1978)
\bibitem[TMN82]{TMN82} J. Thierry-Mieg and Y. Ne'eman, ``Exterior Gauging of an
Internal Supersymmetry and $SU(2/1)$ Quantum Asthenodynamics", in
{\it Proc. of Nat. Acad. of Sciences, USA} {\bf 79} (1982) p.
7068-7072.
\bibitem[TM06]{TM06} J. Thierry-Mieg, ``Chiral Yang-Mills theory, non 
commutative differential geometry, and the need for a Lie superalgebra", 
{\it JHEP {2006} {038}}
\bibitem[Wei67]{Wei67} S. Weinberg, ``A Model of Leptons", {\it Phys. Rev. Lett.} {\bf
19} (1967) 1264.
\bibitem[Yao06]{Yao06} W.-M. Yao {\it et al}, ``Particle Data Group //pdg.lbl.gov" {\em J. Phys.}
{\bf G 33} 1 (2006)
\bibitem[YM54]{YM54} C.N. Yang and R. Mills, ``Conservation
of isotopic spin and isospin gauge invariance," {\em Phys. Rev.}
{\bf 96} 191-195 (1954)

\end{thebibliography}
\end{document}